\documentclass[a4paper,12pt]{article}
\usepackage{cite}
\usepackage{wrapfig}
\usepackage{graphicx}
\usepackage{amssymb}
\usepackage{amsfonts}
\usepackage{amsmath}
\usepackage{longtable}
\usepackage{rotating}
\usepackage{lscape}
\usepackage{epsfig}
\usepackage{multirow}
\usepackage{bm}
\usepackage{lipsum}

\renewcommand{\ln}{\mathop{\rm ln}\nolimits}

\begin{document}
\title{\begin{flushright}
{\small  INR-TH-2023-011}
\end{flushright}
\let\thefootnote\relax\footnotetext{This is a preprint of the Work accepted for publication in
Physics of Particles and Nuclei Letters, ©, copyright 2023, 
Pleiades Publishing, Ltd. }
\,\\
Numerical study of multiparticle production
   in $\lambda\phi^4$ theory}
\author{S.\,Demidov$^{a,b}$,
B.\,Farkhtdinov$^{a,b}$\footnote{E-mail: farkhtdinov@phystech.edu},
 D.\,Levkov$^{a,c}$}
\date{} 
\maketitle

{\small $^{a}$\,Institute for Nuclear Research of the Russian Academy of Sciences, 
  Moscow 117312, Russia
  
$^{b}$\,Moscow Institute of Physics and Technology, 
	Dolgoprudny 141700, Russia
	
$^{c}$\,Institute for Theoretical and Mathematical Physics MSU,
  Moscow 119991, Russia}

\begin{abstract}
  We study multiparticle production in the unbroken
  $(3+1)$--dimensional $\lambda\phi^4$ theory using the semiclassical
  method of singular solutions. We show that the probabilities of these processes are
  exponentially suppressed in terms of a small coupling constant $\lambda \ll 1$
  if the multiplicity of the final state is large:~$n \gg 1$.
  At $ n \ll \lambda^{-1}$ the probabilities agree with well--known 
  perturbative results. At $ n \gg \lambda^{-1}$ they are dominated by loop effects and
  decrease exponentially with $n$, as we show for the first time.
\end{abstract}
\vspace*{6pt}

\noindent
\section*{Introduction}
\label{sec:intro}
We consider multiparticle production in the weakly--coupled 
theory of a real scalar field $\phi(t,{\bm x})$ with the
potential
\begin{equation}
\label{eq:1.0}
V(\phi) = m^2 \phi^2/2+\lambda \phi^4/4\;,
\end{equation} 
where $m^2>0$ is the particle mass and $\lambda \ll 1$ is
the coupling constant.
In the multiparticle processes, 
$n \gg 1$ particles are created in the 
collision of a few particles. The difficulty in
description of these processes appears when $n$
exceeds the inverse coupling constant of the theory $\lambda^{-1}$.

Indeed, the amplitude of producing~$n$ particles at the mass
threshold from one off--shell particle equals~\cite{Brown:1992ay,Voloshin:1992nu}
\begin{equation}
\label{eq:1.1}
{\cal A}_{ n}={\cal A}_{n}^{\rm tree} \left[ 1+B\lambda n^2 + O(\lambda^2 n^4) \right]\, , 
\qquad  {\cal A}_{n}^{\rm tree} = n!\left( \frac{\lambda}{8m^2}\right)^{(n-1)/2}\, ,
\end{equation}
where ${\cal A}_{n}^{\rm tree}$ comes from tree diagrams, 
the term $B\lambda n^2$ with known~\cite{Voloshin:1992nu} numerical constant $B$
represents one--loop correction and multiloop terms
are hidden inside $O(\lambda^2 n^4)$. 
At $\lambda n \ll 1$, the tree contribution is the largest,
and the leading correction comes from one--loop diagrams. 
But at $n \gtrsim \lambda^{-1}$  all contributions are comparable
and the series~\eqref{eq:1.1} break down~\cite{Cornwall:1990hh,Goldberg:1990qk}.
It was argued~\cite{Libanov:1994ug} that the leading parts of multiloop contributions 
at $n\sim O\left(\lambda^{-1}\right)$ can be resummed into the exponent
\begin{equation}
\label{eq:1.2}
{\cal A}_{n} = {\cal A}_{n}^{\rm tree}\; {\rm e}^{F_{\cal A}/\lambda}\, , 
\qquad {\rm where} \qquad F_{\cal A}= B(\lambda n)^2 + O(\lambda^3 n^3) \, .
\end{equation}
However, the function $F_{\cal A}(\lambda n)$ is presently unknown apart
from the leading term of its Taylor expansion at $\lambda n \ll 1$ in Eq.~\eqref{eq:1.2}.

One wonders whether the full multiparticle amplitude~${\cal A}_{n}$ may become
of order one at some $n \sim O\left( \lambda^{-1} \right)$, 
similarly to the tree--level result~${\cal A}_n^{\rm tree}$.
If that is the case, few--particle scattering may lead to a spectacular
multiparticle production at sufficiently high collision
energies $E \sim mn$. 

The exponential form \eqref{eq:1.2} of the amplitude suggests that
multiparticle processes can be described semiclassically
\cite{Khlebnikov:1992af,Diakonov:1993ha,Son:1995wz,Libanov:1997nt}.
The most promising development in this direction is D.T.~Son's
method of singular solutions~\cite{Son:1995wz} which is applicable at
$\lambda \ll 1$ and $n \gg 1$. In this talk we describe its numerical implementation
and demonstrate some preliminary results in the unbroken~$\lambda \phi^4$ theory,
see Ref.~\cite{Demidov:2022ljh} for the follow--up publication. 

\section*{Singular semiclassical solutions -- numerically}
\label{sec:comp_sing_solut}
Following Ref.~\cite{Son:1995wz}, we consider inclusive probability
of transition~${{\rm few}\to n}$ at energy $E$:
\begin{equation}
\label{eq:2.1}
{\cal P}_n ( E)  \equiv \sum_{f} |\langle
f; E,n|\hat{\cal S}\,\hat{\cal O}|0\rangle|^2 \;,
\end{equation}
where the operator $\hat{\cal O}$ creates a few--particle initial 
state, the sum runs over all final states $f$ with fixed energy $E$ and multiplicity $n$,
and $\hat{\cal S}$ is the scattering matrix. To be concrete, we consider the 
in--state operator of the form
\begin{equation}
\label{eq:2.2}
\hat{\cal O} \to \hat{\cal O}_J = \exp 
\left( -\int d^3 {\bm x} \; J({\bm x}) \hat{\phi}(0, {\bm x}) \right) , 
\;\;\; J({\bm x}) = \frac{j_0}{\sqrt{\lambda}} {\rm e}^{- {\bm x}^2/(2\sigma^2)} ,
\end{equation}
that includes a localized classical source $J({\bm x})$ of strength $j_0$ and
width $\sigma$ acting at $t=0$.

\begin{figure}
  \centerline{\includegraphics{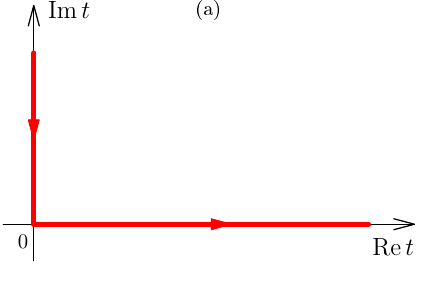}\hspace{8mm}
   \includegraphics{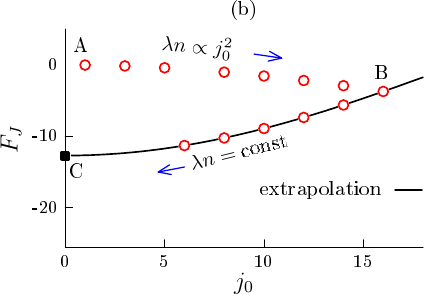}}
   \caption{(a) A contour in the complex time plane for Eq.~\eqref{eq:2.4}.
   Arrows show the direction from past to future. (b) Deformation of the linear 
   solution A into the nonlinear solution B, and then --- into the singular solution C
   with $j_0 = \sigma = 0$. Units with $m=1$ are used.}
   \label{fig:cont_and_extrap}
\end{figure}

The derivation of the classical method~\cite{Son:1995wz} consists of
writing down path integral for Eq.~\eqref{eq:2.1} and evaluating
it in the saddle--point approximation. The result
for the probability is
\begin{equation}
\label{eq:2.3}
{\cal P}_n^{(J)} ( E) \sim {\rm e}^{F_J(\lambda n,\, \varepsilon)/\lambda} \, ,
\end{equation}
where $\varepsilon = E/n-m$ is the mean kinetic energy
of out--particles and we ignore the prefactor. The semiclassical 
exponent $F_J$ is determined using 
the value of the the classical action $S\left[\phi_{\rm cl}\right]$ on the
saddle--point configuration $\phi_{\rm cl}(t,{\bm x})$. 
The latter is complex and satisfies the field equation
\begin{equation}
\label{eq:2.4}
\Box \phi_{\rm cl} + m^2\phi_{\rm cl} + \phi_{\rm cl}^3 = i J({\bm x})\delta (t)\; .
\end{equation}
It is convenient to obtain~$\phi_{\rm cl}$ on
the complex time contour depicted in Fig.~\ref{fig:cont_and_extrap}a.
Boundary conditions for this solution are specified by the initial and
final states of the process: vacuum state at $t \to +i\infty$ and
multiparticle state with fixed~$E$ and $n$ at $t \to +\infty$.
In a nutshell, our semiclassical method consists of solving the above
boundary value problem for $\phi_{\rm cl}$ and calculating $F_J$ using
the solution.

It is worth noting that the operator $\hat{\cal O}_J$ creates $O(J^2/\lambda)$
particles from the vacuum, so that the few--particle initial state is
achieved only at sufficiently small
$J$. As a consequence, the probability~\eqref{eq:2.1} of the
process ${\rm few} \to n$ is restored in the limit
\begin{equation}
\label{eq:2.5}
F(\lambda n, \varepsilon) = \lim_{J \to 0} F_J(\lambda n, \varepsilon) \qquad
{\rm and} \qquad {\cal P}_n ( E) \sim {\rm e}^{F(\lambda n, \varepsilon)/\lambda}\, .
\end{equation}
It was conjectured~\cite{Libanov:1994ug} that the value of $F$ is insensitive to the
details of the few--particle initial state $\hat{\cal O} | 0 \rangle$
and hence independent of the particular form of $J({\bm x})$ in the limit~\eqref{eq:2.5}.
We send $j_0 \to 0$ while keeping $j_0/\sigma = {\rm const}$, which corresponds 
to an infinitesimally weak Gaussian source supported at ${\bm
  x}=0$. One can show that 
the saddle--point configuration $\phi_{\rm cl}$ becomes singular in this limit.

We solve the boundary value problem for Eq.~\eqref{eq:2.4} numerically. 
To this end we adopt units with $m=1$, substitute the spherically--symmetric
Ansatz ${\phi_{\rm cl}=\phi_{\rm cl}(t,r)}$, and introduce a rectangular
spacetime lattice $\{ t_j,r_i \}$ with~$t_j$ covering the complex contour in
Fig.~\ref{fig:cont_and_extrap}a. We use Newton--Raphson method~\cite{NR}
to solve the lattice equations at every accessible value of~$\lambda n$,~$\varepsilon$,~$j_0$,
and~$\sigma$. Once the exponent $F_J$ is found, we extrapolate it to
$j_0 = 0$.

We select physically relevant saddle--point configurations using
the following strategy. At small values of $j_0$~and~$\lambda n \propto j_0^2$
the interaction term in Eq.~\eqref{eq:2.4} can be ignored --- hence the semiclassical 
solution~$\phi_{\rm cl}$ describes creation of particles by 
the source $J$ in the almost free theory. 
The respective solutions of the linearized equation~\eqref{eq:2.4} --- 
unique and definitely physical ---
can be obtained analytically. Using one of them as the first approximation,
we numerically solve the full nonlinear equation and arrive at the true saddle--point 
configurations, see the point A in Fig.~\ref{fig:cont_and_extrap}b. After that
we increase $j_0$~and~$\lambda n \propto j_0^2$ in small steps 
building a continuous branch of numerical solutions,
see the chain of circles~AB in Fig.~\ref{fig:cont_and_extrap}b. 
Once the solution B was found, we consider the limit~\eqref{eq:2.5}. 
To this end, we decrease $j_0$ at {\it fixed}~$\lambda n$,
$\varepsilon$, and $j_0/\sigma$, see the circles between B and
C in Fig.~\ref{fig:cont_and_extrap}b. A polynomial extrapolation of
$F_J$ to $j_0 = 0$ (solid line in Fig.~\ref{fig:cont_and_extrap}b) 
gives the final result for the few--particle exponent~$F(\lambda n,\, \varepsilon)$
(point C).    

\section*{Results}
\label{sec:results}
\begin{figure}
  \centerline{\includegraphics{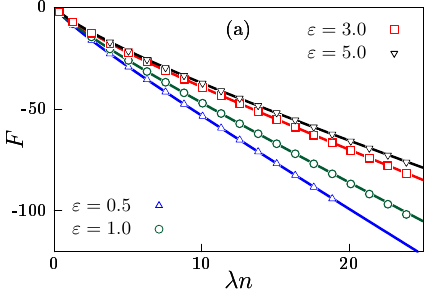}\hspace{10mm}
    \unitlength=1mm
    \begin{picture}(72,50)
      \put(0,0){\includegraphics{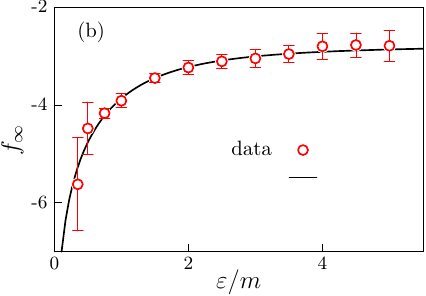}}
      \put(34.5,18.5){\small fit~\eqref{eq:3.4}}
    \end{picture}}
   \caption{(a)~Suppression exponent $F$ as a function of $\lambda n$
   at fixed $\varepsilon/m = 0.5;\;1;\;3;\;5$. Limit
   $j_0 \to 0$ has been already evaluated.
   (b)~The slope~$f_\infty (\varepsilon)$ of the exponent
     at large~$\lambda n$ as a function of $\varepsilon/m$. }
   \label{fig:f_and_df_dn}
\end{figure}

Our numerical data for the suppression exponent $F(\lambda n,\, \varepsilon)$ are
shown by empty circles, squares, and triangles in
Fig.~\ref{fig:f_and_df_dn}a. In the previous  
publication~\cite{Demidov:2021rjp} we demonstrated that
they are close to the well--known tree--level result at~$\lambda n \ll 1$:
\begin{equation}
\label{eq:3.1}
F(\varepsilon, \lambda n) = \lambda n \ln \left( \frac{\lambda n}{16} \right) +
\left[ f(\varepsilon)-1\right] \;\lambda n + O(\lambda n)^2 \qquad {\rm at} \qquad \lambda n \ll 1\, ,
\end{equation}
where the function $f(\varepsilon)$ is known numerically, 
see Refs.~\cite{Bezrukov:1995ta,Bezrukov:1998mei}. Notably, we 
find that in the opposite limit
$\lambda n \gg 1$ the graphs in Fig.~\ref{fig:f_and_df_dn}a are
almost linear:
\begin{equation}
\label{eq:3.2}
F(\varepsilon, \lambda n) = f_\infty (\varepsilon) \lambda n + g_\infty (\varepsilon)
\qquad {\rm at} \qquad \lambda n \gg 1\;.
\end{equation}
In practice, it is convenient to describe all numerical data with the function
\begin{equation}
\label{eq:3.3}
F(\varepsilon,\lambda n) = \lambda n f_\infty (\varepsilon) -
\frac{\lambda n}{2} \ln \left[ \left(\frac{16}{\lambda n}\right)^2
{\rm e}^{2-2f(\varepsilon)+2f_\infty (\varepsilon)} - \frac{2g_\infty(\varepsilon)}{\lambda n}
+1 \right]  
\end{equation}
that interpolates between the asymptotic regimes \eqref{eq:3.1} and \eqref{eq:3.2}.
Fitting numerical results with Eq.~\eqref{eq:3.3},
we obtain solid lines in~Fig.~\ref{fig:f_and_df_dn}a and values
of $f_\infty (\varepsilon)$ and $g_\infty(\varepsilon)$. 
The tilt $f_\infty (\varepsilon)$ of the semiclassical exponent is plotted in
Fig.~\ref{fig:f_and_df_dn}b (circles). 
Notably, it is negative and can be approximated by the function
\begin{equation}
\label{eq:3.4}
f_\infty (\varepsilon) = -\frac{3}{4}\ln \left[ \left( \frac{d_1 m}{\varepsilon}\right)^2 +d_2 \right],
\quad {\rm where} \quad d_1 \approx 11.5\, , \; d_2 \approx 40.2 \, ,
\end{equation}
that has intuitively correct behaviors at $\varepsilon \to 0$ and $\varepsilon \to +\infty$;
see the solid line in Fig.~\ref{fig:f_and_df_dn}b.

In a nutshell, our results imply that multiparticle production is
exponentially suppressed in the unbroken $\lambda \phi^4$ theory at any
$\lambda n$ and $\varepsilon$.

\section*{Conclusion}
\label{sec:concl}
We numerically implemented D.T.~Son's method
of singular solutions in the unbroken $\lambda \phi^4$ theory.
Using this method, we computed the probability of multiparticle production 
${{\rm few} \to n}$. We explicitly demonstrated that the process remains
exponentially suppressed at any energy~$E$ and multiplicity $n$
if the latter is large:~$n \gg 1$.
Moreover, at $\lambda n \gg 1$ the probability
decreases exponentially with $n$. We provided compact fitting expressions
for all of our numerical results.

Numerical calculations were performed on the Computational cluster of 
the Theoretical Division of INR RAS.

\bibliographystyle{pepan}
\bibliography{dubna_proceedings}
\end{document}